\documentclass[conference]{IEEEtran}
\IEEEoverridecommandlockouts

\usepackage{cite}
\usepackage{amsmath,amssymb,amsfonts}
\usepackage{algorithmic}
\usepackage{graphicx}
\usepackage{textcomp}
\usepackage{xcolor}
\usepackage{tabularx}
\usepackage{booktabs}
\usepackage{array}

\newcolumntype{P}[1]{>{\raggedright\arraybackslash}p{#1}}

\def\BibTeX{{\rm B\kern-.05em{\sc i\kern-.025em b}\kern-.08em
    T\kern-.1667em\lower.7ex\hbox{E}\kern-.125emX}}

\begin{document}

\title{Supervised Quantum Machine Learning: A Future Outlook from Qubits to Enterprise Applications}

\author{
\IEEEauthorblockN{Srikanth Thudumu, Jason Fisher}
\IEEEauthorblockA{\textit{ Department of AI} \\
\textit{ Institute of Applied Artificial Intelligence and Robotics (IAAIR)}\\
Germantown, TN, 38139, USA \\
\textit{\{srikanth\}\{jason\}@iaair.ai}}
\and
\IEEEauthorblockN{Hung Du}
\IEEEauthorblockA{\textit{Applied Artificial Intelligence Institute $(A^2I^2)$} \\
\textit{Deakin University}\\
Geelong, VIC 3216, Australia \\
\textit{hung.du@deakin.edu.au}}

}

\maketitle

\begin{abstract} 
Supervised Quantum Machine Learning (QML) represents an intersection of quantum computing and classical machine learning, aiming to use quantum resources to support model training and inference. This paper reviews recent developments in supervised QML, focusing on methods such as variational quantum circuits, quantum neural networks, and quantum kernel methods, along with hybrid quantum-classical workflows. We examine recent experimental studies that show partial indications of quantum advantage and describe current limitations including noise, barren plateaus, scalability issues, and the lack of formal proofs of performance improvement over classical methods. The main contribution is a ten-year outlook (2025-2035) that outlines possible developments in supervised QML, including a roadmap describing conditions under which QML may be used in applied research and enterprise systems over the next decade.
\end{abstract}

\begin{IEEEkeywords} Quantum Machine Learning, Variational Quantum Circuits, Quantum Neural Networks, Quantum Kernel Methods, Hybrid Models, NISQ Devices, Quantum Computing, Error Mitigation, Quantum Federated Learning, Future of Quantum Machine Learning.
\end{IEEEkeywords}

\section{Introduction}
Quantum Machine Learning (QML) has emerged from a cross-fertilization of ideas between quantum computing and classical machine learning. QML aims to utilize quantum computation to improve learning algorithms, with qubits and quantum gates serving roles analogous to neurons and activation functions in classical networks \cite{zeguendry2023quantum, melnikov2023quantum}. This paper focuses on supervised QML, where quantum models are trained on labeled data (either classical or quantum) to perform tasks such as classification and regression. Quantum models operate in exponentially large Hilbert spaces, which in theory provide a broader representational capacity than classical models \cite{huang2021power}. However, leveraging this space effectively requires the development of efficient encoding schemes and circuit architectures that avoid exponential resource overhead, as current quantum systems cannot fully exploit this space due to hardware constraints \cite{hu2024overcoming, garcia2024mitigating}.

Theoretical results indicate that certain tasks can be learned more efficiently by quantum models, and exponential improvements in sample complexity have been demonstrated when learning from quantum data compared to classical data \cite{huang2021power, huang2022quantum, molteni2024exponential}. These findings suggest that QML may outperform classical approaches under specific conditions. However, current quantum hardware presents significant limitations \cite{upadhyay2024trustworthy, huang2022quantum}. Noisy Intermediate-Scale Quantum (NISQ) devices have limited qubit counts and coherence times, which constrain the size and depth of quantum circuits that can be executed reliably \cite{hu2024overcoming, garcia2024mitigating, mathur2025federated}. As a result, most QML implementations today follow hybrid quantum-classical strategies: a classical computer optimizes the parameters of a quantum circuit (often through iterative feedback), keeping quantum processing short enough to mitigate noise \cite{zaman2024comparative, cho2024machine, nicoli2023physics}. As of 2025, practical demonstrations of quantum advantage in supervised learning remain limited \cite{hibat2024framework, thanasilp2024exponential}. Existing results are primarily confined to synthetic benchmarks or small-scale datasets, and a reproducible advantage over classical methods on real-world tasks has yet to be empirically demonstrated \cite{upadhyay2024trustworthy, huang2022quantum}. 

This paper presents an overview of the current state of supervised QML. We review existing techniques and recent developments that have been empirically validated and are practically implementable under Noisy Intermediate-Scale Quantum (NISQ) constraints, including experimental studies and reported claims of quantum advantage. We analyze key challenges that hinder progress, such as noise, trainability issues (e.g., barren plateaus), scalability limitations, and data availability, along with theoretical gaps, including the lack of general proofs for quantum advantage. Finally, we outline future research directions, such as error mitigation, quantum data generation, federated learning, and quantum AutoML, and provide a perspective on anticipated developments over the next five to ten years. This review draws on recent literature from peer-reviewed journals and conferences, encompassing both theoretical and experimental research in QML.

\section{Key Techniques in Supervised QML}

\begin{figure*}[!ht]
    \centering
    \includegraphics[width=\linewidth]{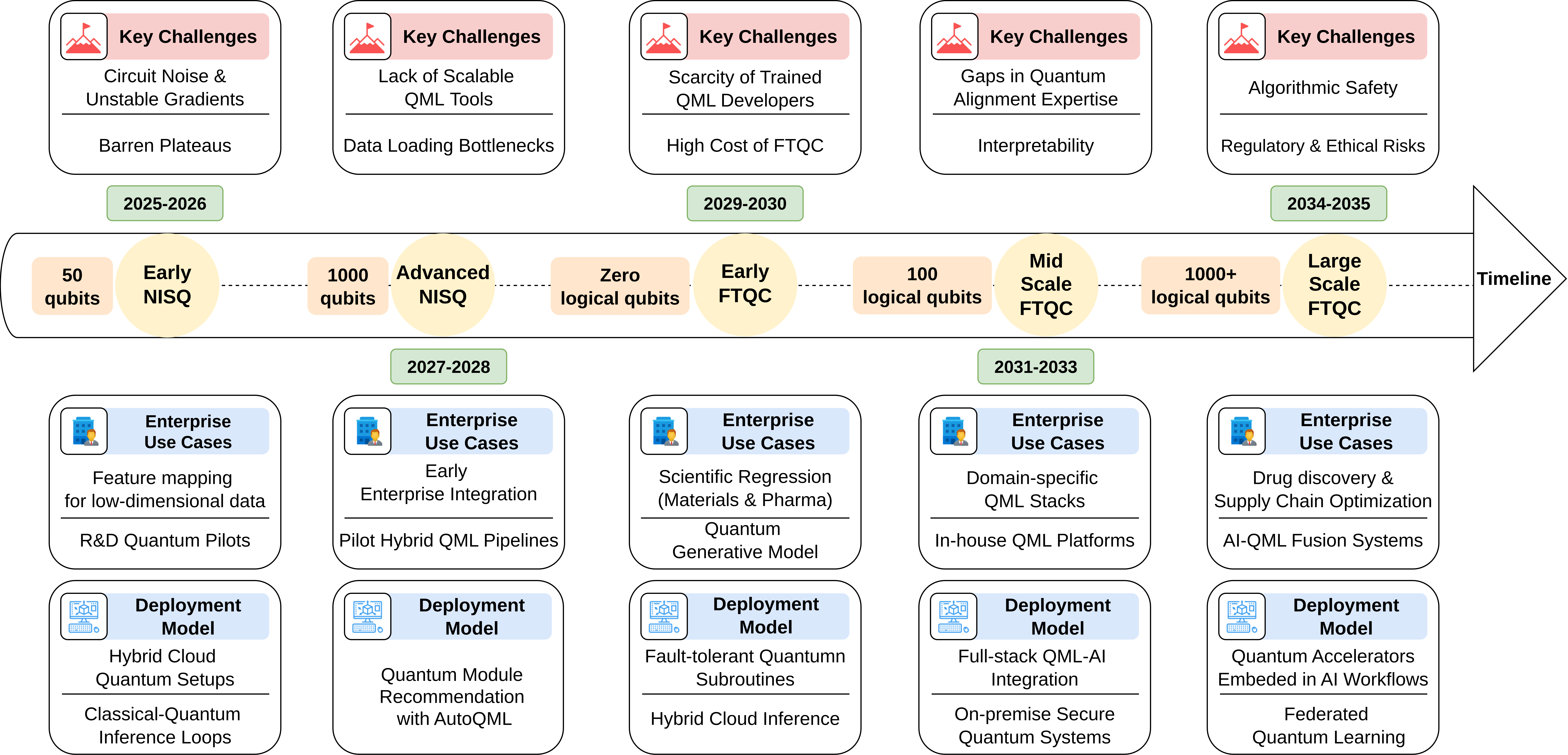}
    \caption{Projected Roadmap for Supervised QML (2025–2035).}
    \label{fig:overview}
\end{figure*}

This study focuses on techniques that have been empirically validated and are practically implementable in supervised QML under Noisy Intermediate-Scale Quantum (NISQ) constraints. Recent studies have demonstrated their strong model capacity \cite{chen2024evolutionary}, benchmarked effectiveness in quantum kernel learning \cite{schnabel2025quantum}, and the critical role of encoding strategies in influencing quantum performance \cite{rath2024quantum}. Moreover, hybrid workflows remain the most feasible path to deployment, balancing quantum advantages with classical reliability \cite{cranganore2024paving}.

\textbf{Variational Quantum Circuits (VQCs)}, often referred to as quantum neural networks (QNNs), employ parameterized quantum gates that are optimized via classical algorithms to learn from data \cite{bharti2022noisy, caro2022generalization}. Architectures such as quantum convolutional neural networks (QCNNs) have shown effective performance in tasks like quantum phase recognition using shallow circuits \cite{panadero2024regressions}. Recent research indicates that deeper data re-uploading architectures can represent complex functions using fewer qubits compared to simple linear models, positioning QNNs as promising candidates for near-term applications \cite{kolle2024study}.

\textbf{Quantum-Enhanced Kernel Methods} embed classical data into high-dimensional quantum states, enabling linear classifiers such as support vector machines (SVMs) to separate complex classes \cite{franco2025quantum}. These quantum kernel classifiers have been tested on real quantum hardware and have achieved competitive classification accuracy despite noise \cite{thanasilp2024exponential}. However, challenges such as kernel concentration due to noise and entanglement complexity must be addressed to scale these methods to larger systems \cite{schetakis2025data}.

\textbf{Quantum Feature Mapping} is a foundational component in QML. Common encoding strategies include basis, amplitude, and angle encoding. Each technique presents trade-offs among qubit requirements, expressive power, and trainability. Data re-uploading, which involves feeding input features into the circuit multiple times, has been shown to enhance learning performance and alleviate training difficulties \cite{souza2024regression}.

\textbf{Hybrid Quantum-Classical Workflows} are currently the most prevalent design in supervised QML \cite{wang2024hybrid, deluca2021survey}. These systems combine classical pre- or post-processing with quantum computation. Hybrid models, including classical-quantum neural networks, have demonstrated robustness to noise and have shown modest performance advantages over classical counterparts in high-noise settings \cite{schetakis2025data}.

\section{Latest Advancements and Empirical Studies}

Despite hardware limitations, recent years have witnessed significant experimental progress in supervised QML, with researchers pushing the boundaries of what can be achieved on actual quantum devices. This section highlights landmark results from 2021 to 2024 that illuminate the current state of the art.

\subsection{Proof-of-Concept Quantum Advantage Demonstrations}

Huang et al. \cite{huang2022quantum} demonstrated substantial quantum advantages in learning tasks. In their study, a quantum computer learned properties of physical systems (such as predicting outcomes of measurements) using exponentially fewer experiments than a classical approach would require.

This research work, published in \textit{Science}, further underscores a theme: quantum advantages in the near term may be most achievable for quantum-native data, rather than classical datasets like images or text.
In \cite{huang2021power}, the authors showed that many classically hard functions can be easily solved by classical neural networks given enough data. They did, however, identify and experimentally validate an engineered dataset where a quantum model outperformed classical models in prediction error.

\subsection{Scaling Up QML Experiments}
In 2024, researchers at IonQ reported what is considered the largest quantum classification task demonstrated to date \cite{widdows2024quantum}. The experiment involved classifying over 10,000 textual data points in a natural language processing (NLP) context. The team, led by Dominic Widdows, implemented a topic classification task using a trapped-ion quantum computer. The goal was to explore whether shallow quantum circuits could process text representations in a way that yields meaningful learning performance on classification tasks. The workflow integrated classical word embeddings with quantum circuits acting as kernel approximators. By encoding these embeddings into quantum states and measuring their inner products, the system performed classification based on quantum-enhanced kernel methods. The researchers achieved an average accuracy of 62\% on a five-way classification task, demonstrating that some NLP workloads can be executed on current quantum hardware with statistically significant outcomes \cite{widdows2024quantum}. This work provides a notable example of applying hybrid quantum-classical techniques in the NISQ era. It suggests that quantum NLP, although in its infancy, may evolve to complement classical models for specific high-dimensional tasks where kernel-based learning is advantageous.

\subsection{Generalization and Training Advances}
In this section, we aim to highlight theoretical results that provide insight into when and how quantum models can generalize from limited data, despite hardware constraints. At the same time, we recognize that trainability, especially under realistic noise models, remains a bottleneck. This contrast with some of the assumptions in classical learning theory. 

Recent theoretical work has begun to clarify how and when QML models can generalize from training data. Caro et al. \cite{caro2022generalization} derived bounds showing that the generalization error of a QML model scales approximately as $\sqrt{T/N}$, where $T$ is the number of trainable gates and $N$ is the number of training examples. Importantly, when only a subset $K \ll T$ of these parameters are significantly updated during training, the bound improves to $\sqrt{K/N}$. These findings suggest that quantum models may be able to generalize effectively even when full-parameter training is infeasible, which is particularly relevant in the Noisy Intermediate-Scale Quantum (NISQ) era.

However, improved generalization potential does not imply that training quantum models is straightforward. Trainability remains one of the most critical challenges in supervised QML. Anschuetz and Kiani \cite{anschuetz2022beyond} demonstrated that, even in the absence of barren plateaus, many shallow variational circuits can be untrainable due to the presence of poor optimization landscapes, including numerous local minima. Moreover, when noise is present, the number of samples required for successful optimization may grow super-exponentially, posing significant scalability barriers for practical applications.

These results establish a promising theoretical foundation for understanding generalization in QML; however, they also underscore key differences from classical learning systems. Specifically, the optimization dynamics and error landscapes in quantum models tend to be less predictable and more susceptible to hardware-induced noise and variability. Consequently, further progress in training methodologies, error mitigation techniques, and quantum circuit design is essential for advancing the practical applicability of supervised QML.

\section{Key Challenges}

While supervised quantum machine learning (QML) holds theoretical promise, several critical obstacles must be overcome before practical quantum advantage becomes a reality. These challenges span hardware constraints, training instability, scalability issues, and foundational theoretical gaps.

\subsection{Noisy Hardware and Circuit Depth Constraints}

Current quantum devices, especially those in the Noisy Intermediate-Scale Quantum (NISQ) era, suffer from gate errors, decoherence, and imprecise readouts \cite{cerezo2021variational, georgopoulos2021modelling}. These limitations drastically restrict the circuit depth and qubit count that can be reliably executed. As circuits become deeper, the cumulative noise often overwhelms the signal. Although mitigation strategies like zero-noise extrapolation and probabilistic error cancellation exist, they increase the number of required samples and demand precise calibration \cite{hu2024overcoming, garcia2024mitigating}. Until large-scale fault-tolerant quantum computers become available, meaningful QML applications will likely remain constrained to shallow circuits on small-scale datasets.

\subsection{Barren Plateaus and Training Instability}

One of the most significant barriers in training quantum models is the barren plateau phenomenon, where gradients vanish exponentially with the number of qubits or circuit layers \cite{mcclean2018barren}. This causes the model to receive little to no training signal, making gradient-based optimization ineffective. Even in settings without barren plateaus, variational circuits can suffer from poor local minima and flat regions that hinder learning \cite{anschuetz2022beyond}. Gradient-free methods and problem-specific circuit designs offer some relief, but a general solution for training stability remains an open problem.

\subsection{Scalability with Data and Model Size}

Unlike classical machine learning, which scales favorably with large datasets, QML faces unique scalability challenges. Encoding classical data into quantum states often incurs significant overhead. Amplitude encoding, for example, can require complex subroutines and may not be efficient for high-dimensional data \cite{wang2021understanding}. Moreover, quantum circuit evaluations are computationally expensive, limiting the size and number of training examples that can be feasibly processed. This leads many QML studies to rely on downsampled or synthetic datasets, which reduces their relevance for real-world applications.

\subsection{Lack of Standardized Benchmarks}

Progress in classical machine learning has been driven by standardized datasets and benchmarks like ImageNet and GLUE. In contrast, QML lacks a shared set of benchmark tasks that can meaningfully demonstrate quantum advantage. Most QML studies use small, synthetic datasets that may not reflect practical use cases. While efforts like QDataSet are a start \cite{perrier2021qdataset}, the field needs community-driven initiatives to define quantum-relevant benchmarks for classification, regression, and generative modeling.

\subsection{Unresolved Theoretical Foundations}

Many QML models operate without formal guarantees of quantum advantage. In some cases, it has been shown that quantum models are classically simulable, which undermines their expected benefit \cite{huang2021information, ciliberto2018quantum}. Additionally, existing proofs of quantum advantage often rely on oracle models or contrived problems that don't map directly to real-world tasks. As a result, it remains unclear which learning problems genuinely benefit from quantum resources and which are more efficiently solved using classical methods. Without a solid theoretical foundation, the long-term value of QML remains speculative.

\section{Future Outlook}
Looking forward 5–10 years, supervised QML is poised at an interesting juncture: there is both cautious realism and optimism in the community. Here we outline promising directions and our perspective on the future trajectory of QML, grounded in expert views and recent trend analysis (refer to Figure \ref{fig:overview}). 

\subsection{Error Mitigation and Towards Fault-Tolerance} In the near term, substantial effort will focus on improving the effective performance of QML models through advanced error mitigation techniques. Methods such as zero-noise extrapolation, virtual distillation, and learning-based error mitigation may be game-changers for QML. For example, one could imagine training a variational model with an objective function that explicitly penalizes measurement noise or using an auxiliary quantum error detection code in the loop to filter out corrupted runs. These techniques won’t eliminate noise but could extend the scale at which QML models operate usefully by, say, a factor of 2–3 in circuit depth or qubit count. On the hardware side, the gradual increase in qubit numbers (IBM and others are now in the 100–400 qubit range, aiming for 1000+ in a couple of years) combined with better gate fidelities will slowly reduce the gap between what’s implementable and what’s algorithmically desirable. If error rates drop by an order of magnitude, we could start to see shallow-depth quantum circuits with meaningful size (50–100 qubits) reliably implementing QML models, which might be enough to attempt non-trivial datasets. 

The ultimate goal is Fault-Tolerant Quantum Computing (FTQC), which is required for robust advantages. Fault-tolerance would allow deep quantum circuits analogous to deep neural networks, unlocking algorithms that are currently out of reach. A fault-tolerant quantum computer could, for instance, implement the quantum analog of a deep convolutional network or a Transformer-like model on very large data, or run Grover-type search subroutines to find optimal model parameters faster than classical training could. Most experts temper this with timeline caution: fault-tolerance is likely a decade or more away. Until then, early fault-tolerant techniques (perhaps using a few error-corrected logical qubits combined with many physical qubits in raw mode) might start appearing. 
As outlined in Figure \ref{fig:overview}, during the 2025–2028 period, quantum models will remain within the Early and Advanced NISQ phase, where techniques such as hybrid cloud quantum setups and classical-quantum inference loops will dominate, requiring robust error mitigation strategies to compensate for circuit noise and barren plateaus. By 10 years, we are likely to see at least small-scale fault-tolerant QML demonstrations. The consensus is that truly transformative QML applications – ones that outperform classical ML on real-world big data problems – will coincide with the advent of fault-tolerant hardware. But there is a lot of impactful science to be done along the way. 

\subsection{Benchmark Applications and Quantum Advantage Candidates} One of the most important developments we anticipate is the crystallization of benchmark QML applications. Likely areas include: quantum chemistry and materials (where the data is quantum mechanical and quantum models naturally fit), optimization problems (learning heuristics for NP-hard problems where quantum effects might help explore solutions), finance (where there is appetite for any edge and data might be low signal-to-noise), and certain scientific data analysis tasks (e.g. high-energy physics where data are enormous in volume but perhaps compressible by quantum feature maps). In the next few years, we expect community-driven efforts to create standard datasets for QML. For instance, a dataset of molecular spectra classification where the input is given as quantum states, or a “Quantum ImageNet” consisting of quantum states with certain properties to be classified. These will serve as playgrounds for QML algorithms and help identify what works best. As these benchmarks emerge, we may see the first clear instances of quantum models beating classical ones by a margin in practical metrics (accuracy and efficiency). A plausible early example might be a hybrid quantum kernel method that classifies phases of matter or detects anomalies in quantum sensor data, succeeding with far fewer experiments than any classical analysis, essentially a continuation of the quantum-native advantage line of work. Another candidate is in the field of metamodeling or scientific regression: using QML to learn the behavior of a physical system (like climate or fluid dynamics) more efficiently than classical models, by encoding differential equations into quantum operators.

\subsection{Quantum Data Generation and Synthetic Data} An intriguing emerging area is using quantum computers to generate quality data for machine learning. With random quantum circuits, quantum systems can produce complex high-dimensional data distributions. This could lead to the development of quantum generative models (QGMs) (e.g., Quantum Circuit Born Machines and Quantum Generative Adversarial Networks) that can be trained to synthesize high-quality data given a target distribution. These quantum models could further address the data scarcity issue by augmenting training sets with quantum-generated samples. For instance, QGMs could learn to produce realistic yet hard-to-simulate examples of a phenomenon, which a classical model would struggle to imitate. We observed a hint of this in the IonQ NLP work \cite{widdows2024quantum}. In the future, QGMs might be used to create novel molecular structures with desired properties, effectively doing creative design in the quantum feature space that classical generative models cannot explore. Additionally, by exploiting quantum superposition, one could potentially amplify the occurrence of rare but critical patterns or events during training. This would be more efficient than classical Monte Carlo sampling techniques. 

\subsection{Privacy and Federated Learning} As QML matures, practical considerations like data privacy and distributed learning will come into play. Quantum Federated Learning (QFL) is an evolving concept where multiple quantum computers (or quantum and classical nodes combined) collaboratively train a model without sharing raw data, akin to classical federated learning in sensitive domains like healthcare. Preliminary studies have proposed frameworks for QFL, highlighting how a quantum model’s parameters could be shared and aggregated while each quantum node sees only its local data \cite{li2025quantum, song2024quantum, zhang2022federated}. This could be particularly useful if quantum computers are deployed at different data centres or institutions (for example, hospitals each with a small quantum co-processor handling their private data). QFL could leverage quantum secure communication for parameter updates, making the federated scheme end-to-end quantum secure. Another aspect is quantum differential privacy, ensuring that quantum models do not inadvertently leak information about training data. Research in 2023 \cite{watkins2023quantum} showed initial algorithms for training quantum models with differential privacy guarantees, which will be crucial if QML is used on personal medical or financial data. These developments ensure that as we push performance, we also keep an eye on ethical and legal considerations surrounding ML. 

\subsection{Automated Quantum ML (AutoQML)} In classical ML, AutoML techniques automate the design and tuning of models. We foresee analogous efforts in QML to automate finding the best quantum circuit ansatz, hyperparameters, or even whether to use a quantum or classical model for a subtask. A near-term example is using classical AI to search for efficient variational circuit structures (number of qubits, connectivity, gate types) that optimise a given validation score. Early versions of this are appearing as researchers use reinforcement learning agents to propose circuit designs or as they borrow ideas like Neural Architecture Search for quantum circuits. In the long run, AutoQML could extend to selecting optimal data encodings or splitting workloads between classical and quantum parts in a pipeline automatically. A compelling vision is a pipeline where, given a dataset, an AutoQML framework tries various combinations of classical and quantum models and delivers the best composite solution perhaps finding that certain data features are best handled by a small quantum kernel model while others are tackled by a classical neural net, for example. Such tools would greatly lower the barrier to entry for non-experts to utilize QML, abstracting away the trial-and-error in designing quantum circuits. Given the complexity of the QML design space, automation might even be necessary to discover non-intuitive high-performing hybrid architectures that humans might not think of. 

\subsection{Timelines and Expected Impact} In terms of timelines, the next 5 years will likely see incremental but important advances: demonstration of quantum advantage on at least one or two specialized supervised learning problems (probably involving quantum data or extremely low-data regimes), quantum hardware reaching ~1000 qubits with error rates around or below $10^{-3}$, and the emergence of integrated quantum-classical cloud platforms where QML algorithms can be deployed similarly to classical ML pipelines. We also expect more cross-disciplinary collaboration, for instance, quantum computing experts working with domain scientists in chemistry or medicine to apply QML to pressing problems, thereby discovering new algorithmic requirements and innovations. The 10-year horizon might bring the early fault-tolerant machines into play, maybe tens of logical qubits that allow deeper circuits. At that stage, if QML research stays on course, we could witness genuinely useful applications: e.g., a quantum-enhanced drug discovery pipeline that identifies candidate molecules faster than any classical method, or a quantum recommendation system that captures subtle correlations in user data that classical models miss. It’s worth noting that there is also a possibility that certain aspects of QML get realized in analogue quantum systems (quantum annealers or optical processors) sooner than digital gate-based systems, which could provide intermediate practical benefits. Regardless of the exact path, a realistic consensus is that quantum machine learning will not replace classical ML broadly in the near-term; instead it will become a specialized tool in the ML toolbox, used when quantum resources offer a clear edge. The hype that quantum computers will revolutionize all of AI is being replaced by a nuanced view: they will likely revolutionise specific subdomains of AI where either the data or the computation aligns with quantum advantages. In other areas, classical ML (which itself keeps improving) may remain superior. In concluding this outlook, it’s also important to manage expectations. The road to quantum advantage in machine learning is long and winding. We may encounter a “quantum winter” if progress is slower than hoped and funding enthusiasm wanes, similar to past AI winters. But the field has matured substantially, and even partial successes (like the quantum advantage for learning physics experiments) fuel continued investment. The interplay of theory and experiment will be crucial: theory to pinpoint where quantum should help, and experiment to verify and inspire new theory when results deviate from expectations. We also note the human factor, developing talent fluent in both quantum computing and ML is essential. Interdisciplinary training programs and collaborations will continue to grow, supporting long-term development in the field.

\section{Conclusion}
Supervised quantum machine learning (QML) applies quantum computing methods to tasks in classical machine learning. This paper reviewed key techniques, recent experiments, and current challenges, including hardware noise, optimization issues, and limited benchmarks. The main contribution is a ten-year outlook from NISQ to fault-tolerant quantum computing, with a roadmap linking hardware development to use cases, deployment models, and technical constraints.  Near-term impacts are expected in niche domains, with broader deployment dependent on the development of fault-tolerant quantum machines. The next decade will be crucial, requiring patience and sustained research investment. Importantly, the exploration of QML is already enriching both quantum computing and classical ML, advancing our understanding of learning as a physical process. With cautious optimism, the field moves toward a future where quantum-enhanced AI becomes an integral part of the machine learning ecosystem.

\bibliographystyle{IEEEtran}
\bibliography{references}

\end{document}